# Anomaly Detection Based on Access Behavior and Document Rank Algorithm

**Prajwal R Thakare, M.Tech**
*IT Dept, ASTRA, Bandlaguda,*

**K. Hanumantha Rao, M.Tech**
*Associate Professor CSE Dept, ASTRA*

**Abstract:-**Distributed denial of service (DDoS) attack is ongoing dangerous threat to the Internet. Commonly, DDoS attacks are carried out at the network layer, e.g., SYN flooding, ICMP flooding and UDP flooding, which are called DDoS attacks. The intention of these DDoS attacks is to utilize the network bandwidth and deny service to authorize users of the victim systems. Obtain from the low layers, new application-layer-based DDoS attacks utilizing authorize HTTP requests to overload victim resources are more undetectable. When these are taking place during crowd events of any popular website, this is the case is very serious. The state-of-art approaches cannot handle the situation where there is no considerable deviation between the normal and the attacker's activity. The page rank and proximity graph representation of online web accesses takes much time in practice. There should be less computational complexity, than of proximity graph search. Hence proposing Web Access Table mechanism to hold the data such as 'who accessed what and how many times, and their rank on average" to find the anomalous web access behavior. The system takes less computational complexity and may produce considerable time complexity.

**Keywords: -** Anomaly, DDos- Distributed denial of service, SYN-Flooding, HTTP Requests.

## 1. INTRODUCTION

Network security consists of the provisions and policies adopted by a network administrator to prevent and monitor unauthorized access to the network, misuse, modification, or denial of a computer network and network-accessible resources. Network security involves the authentication of access to the data in a network, which is managed by the network administrator.

To access information and programs inside their authorization users have to chose an ID and password. Network security surrounds a variety of computer networks, both private and public; those are used in everyday jobs organizing transactions and communications between government agencies, individuals and businesses. Network can be divided into private and the other access by all. Network security is involved in government organizations, enterprises, and other types of private institutions. By assigning a unique name and a corresponding password is the most common way of protecting a network resource.

In the past, hackers were a highly skilled and knowledgeable programmer who knows the details of computer networks and communications and how to utilize vulnerabilities. Now a day's hacking tools are easily available on internet so, anyone can download the tools to hack anything from the Internet.

Anomaly detection is the technique to find where the behavior is different than normal behavior. The behavior, which are detected are called anomalies. Anomalies are referred to as outliers, change, variation, surprise, aberrant, intrusion, anomaly, etc.

There are three categories of anomaly detection techniques available; autonomously anomaly detection techniques are to detect anomalies in an unlabeled test data set. The Supervised anomaly detection techniques require training as classifier as well as data set that are labeled as "normal" and "abnormal". Semi-supervised anomaly detection techniques is constructed a model, representing normal training data set for a normal behavior, and testing the occurrence of a test instance to be generated by the determination model.

## 2. LITERATURE SURVEY

2.1 Graph-based Rare Category Detection

In this paper we had already done graph-based technique for rare category detection and name Graph-based Rare Category Detection (GRADE). It uses a global similarity matrix which is inspired by manifold ranking algorithm, that gives us a more compact result in cluster for minority classes; where the density changes by selecting the examples from the region, it eliminates the acceptance that the majority classes and the minority classes are divisible. In the case, when the information about the data set is suppose to be not present, we have developed a reorganize version of GRADE known GRADE-LI, this algorithm only needs an input, as upper bound on the portion of all the minority classes. Both GRADE and GRADE-LI algorithms also work with graph data.

Graph-based Rare Category Detection (GRADE) is one of the rare detection methods. This results in sharp changes in local density near the boundary of the minority classes and thus makes it easier to discover those classes. Furthermore, we modify the GRADE algorithm to get the GRADE-LI algorithm, which requires less information compared with the GRADE algorithm, and thus is more suitable for real applications.

GRADE-LI algorithm takes less information about data set than GRADE and it can be achieve better performance in most cases. Furthermore, our algorithms can deal with both data with feature representation and graph data, where as existing rare category detection methods can only work with data with feature representation. Compared with GRADE, GRADE-LI is more suited for real applications.

2.2 Efficient Algorithms for Mining Outliers from Large Data Sets

Distance-based outlier is based on the distance of a point from its $k^{th}$ nearest neighbor. Each point is ranked depending upon the distance of it from the $k^{th}$ nearest neighbor and declares the top n points in this ranking to be anomalous or outliers. Based on the nested loop join and index join algorithms, for mining outliers, we have develop a highly efficient partition-based algorithm. This algorithm first create partitions or divide the input data set into dislocate subgroups, and then prune full partitions as soon as it is purposeful that they cannot have outliers. This results in considerable savings in computation.

However, much of this recent work has focused on finding "large patterns." By the phrase "large patterns", we mean characteristics of the input data that are exhibited by a significant portion of the data.

*Partition Based:*

Partition-based algorithm is to first partition the data space by using clustering algorithm, and then reduce partitions as soon as it can be determined that they cannot contain outliers. We briefly describe the steps performed by the partition based algorithm below, and defer the presentation of details to subsequent sections.

2.3 Anomaly Detection Using Proximity Graph and PageRank Algorithm

Anomaly detection techniques are widely used in a variety of applications, e.g., security systems, computer networks etc.

In this paper analyzes an approach anomaly detection using proximity graph and page rank algorithm. In that we apply PageRank algorithm with the proximity graph vertices comprised of data points as vertices, which produces a score determining the extent to which each data point is anomalous. In previous work first forming a density estimate by using training data. In this case the page rank algorithm produces point wise density estimation for the data points in an asymptotic sense, and with much less estimation effort. As a result, big developments in terms of running time are witnessed while maintaining similar detection performance. Experiments with fabricated and real-world data sets illustrate that the proposed approach is computationally tractable and scales well to large high-dimensional data sets.

Anomaly detection is also known as outlier it refers to the problem of discovering problem from a given data set that do not confirm to some normal behavior. Anomaly detection techniques are applied in variety of domain including financial turbulence detection, network monitoring, credit card fraud prevention and system intrusion discovery. In this paper we propose unsupervised anomaly detection scheme using proximity graph and PageRank algorithm. Our algorithm determines the ranking of which are the most anomalous by taking input as a unlabeled data. From data measurements we construct a proximity graph; it indicates similar data points with one node of each data point and edges between nodes. We then audit the stationary distribution of a random walk on this graph, following a difference of the popular PageRank algorithm.

*Proximity Graph:*

Proximity graphs are widely used in machine learning, e.g., for clustering, manifold learning and semi supervised learning. The proximity graph is use for observation of the user it observe the behavior of each user who accesses the web pages.

*PageRank Algorithm:*

The PageRank algorithm is first introduced by Page and Brin, and employed by Google to rank web pages. PageRank is closely associated to random walks. Depends on the graph of the web PageRank determine the ranking for every web page. PageRank has in many applications such as search, browsing, and traffic estimation. Each and every page has number of backlink and forward link. We cannot know whether we can establish all the backlinks of a specific web page, and then we come to know its entire forward links at that time.

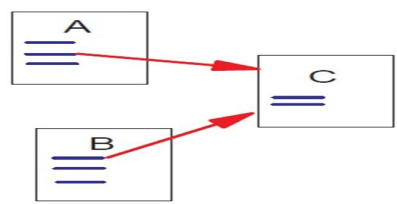

Fig 1: A and B are backlink of C

According to number of backlinks web pages vary considerably. Each URL is converted into a unique integer, and those converted unique integers are storing each hyperlink using integer ID in a database to identify pages. Memory is allocated for the weights for every page.

## 3. SYSTEM ANALYSIS

Firewalls have simple rules such as to allow or deny protocols, IP addresses or ports. Some DoS attacks are too complex for today's firewalls, e.g. if there is attack on the port 80 (web service), firewalls cannot intercept that attack because they cannot distinguish good traffic from DoS attack traffic. However, firewalls can effectively restrict users from launching simple type attack such as flooding from machines beyond the firewall. Flooding is a DDoS attack that is originates to carry a service down by flooding it with huge amounts of traffic. Flooding attacks happen when a service or network becomes so overload with packets commencing defective connection requests that it can no extensive process a genuine connection requests. By flooding attack a server or host with connections that cannot be accomplished by flooding attack, the flooding attack eventually fills the host's memory buffer. Once this buffer is full, then no further connections can be made, and then result is a Denial of Service. Indefinitely interrupt or suspend services of a host connected to the Internet.

In general terms, the DoS attacks has been implemented by either forcing the targeted computer(s) to reset, or consuming its resources so that it can no longer provide its intended service or obstructing the communication media between the intended users and the victim so that they can no longer communicate adequately.

The SYN packet is sent to indicate that a new connection is to be established. The ID for the session is the number that is stored in the **SID** field of the SYN packet.

An outlier is an observation point that is distant from other considerations. An outlier may be due to variability in the measurement or it may indicate experimental error; the latter are sometimes excluded from the data set

A dataset (or data set) is a collection of data.

A data point is measurement of a single member from the group of the member. In a study of the source of money demand with the unit of observation being the independent, a data point might be the values of wealth, income, number of reliant, age of individual. Statistical deduction about the population would be managed using a statistical sample including of dissimilar such data points.

The Cluster analysis itself is not one exact algorithm, but the common task to be solved. It can be concluded by various algorithms that differ somewhat in their notion of what creates a cluster and how to efficiently find them.

A SYN flood is a form of DOS attack in which an attacker sends a succession of SYN requests to a target's system in an attempt to absorb enough server resources to cause the system unresponsive to legitimate traffic.

## 4. SYSTEM DESIGN
4.1 Phases of the work
- Because the proposed system is a machine learning approach,
- It works on
    - Training Phase
    - Testing Phase

4.1  Training phase

4.1.1.1 Access Log Parsing
- The accesses of all users are stored in the access log file.
- The access log files cannot be used for direct comparison
- The access log file is preprocessed to analyze Client IP, Request, and Referrer from each user access log.

4.1.1.2 Web Access Table (WAT)
- This module analyzes the access frequency for each document
- It can be calculated as:

    = (No.Of.Hits for a page per user)/ (Total Number    of Logs) The value constantly in between 0 to 1.
- Training time WAT represents the standard user access behavior.

4.2.1   Testing Phase

4.2.1.1 User Request Access
- The module analyze the user requested (URI).
- It also analyzes the referrer URI.
- The user profile is stored for more processing.

4.2.2.2 Document Matrix
- For every fixed intermission of time, the user-profiles are processed for calculating the Document Matrix.

- Each individual user Document Matrix prepared.

    4.2.2.3 Anomaly Detection

- User WAT is cross compared with the training time of WAT.
- If any URI crosses or under outflow the Training Time WAT of a predefined threshold.
- That user is considered as an anomalous user.
- Those anomalous users are reported to the administrator.

    4.2.2.4 Administration Interface

- This system monitors the anomalous activity.
- The anomalous behavior of any user is analyzed to administrator.
- Administrator allows login, view the anomalous activity.

## 5. SYSTEM ARCHITECTURE

A System Architecture is the conceptual model that defines the structure, behavior, and more views of a system, organized in a way that supports reasoning about the structures of the system.

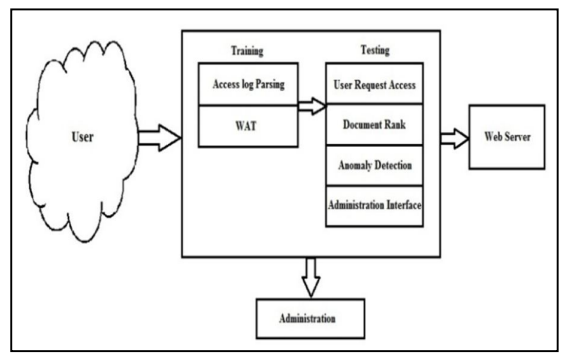

Fig 2: System Architecture

An attacker send the request to server to get resources of any websites, in our system two phases are there training and testing.

In training phase the user accesses are stored in the access log file, that files cannot be used for direct comparison, that file is preprocessed to identify Client IP, Request, referrer from each user access log. After that Web Access Table (WAT) identify the access frequency for each document, training time WAT represents the standard user access behavior.

In testing phase user request access identifies the user requested, it also identifies the referrer URI, the user profile is stored for further processing. For every fixed interval of time, the user-profiles are processed for calculating the Document Rank User WAT is cross compared with the training time WAT. If any URI crosses or under flow the Training Time WAT for a predefined threshold, that user is treated as an anomalous user. Administrator interface monitors the anomalous activity he anomalous behavior of any user is reported to administrator.

## 6. IMPLIMENTATION

HTTrack allows users to download World Wide Web sites from the Internet to a local computer. By default, HTTrack groups the downloaded site by the original site's relative link-structure.

By using HTTRACK we are downloading web WWW sites from the Internet to our local system that we are going to attack.

Get All Request, we will get all requests of the users, it will get IP address, request and time of the user who accessed web sites and how many times Perform Training, we are applying training on that accessed file, it will open access log file and store the observable IP, request and timestamp in to trained.dat file.

Attack on web Document, open the file, in that file url is present that we are going to attack.

Perform Testing, applying testing on that accessed file, it will open access log file and store the observable IP, request in to file.

Calculate Difference, it will open file. Then testing is created and verify with am_test.dat. Then it will calculate document rank and show the result to our application.

## 7. CONCLUSION

In this work, we propose a framework for anomaly detection using Access Behavior and Document rank algorithm. This is an unsupervised, nonparametric, solid estimation free approach, readily extending to high dimensions. Various parameter selection, time complexity guarantees and possible extensions are discussed and investigated. We see several possible directions for future development.

Another direction is to make the framework online. At this stage, our algorithm operates in a batch mode. Given a set of surveys, after announcing the potential anomalies once, the algorithm terminates. However, in practice, it is quite common for successive measurements to come incrementally as time passes by. Once a new observation is available, we do not want to run the whole algorithm from start again. The time complexity of our framework has already been shown to be $O(n^2)$, which is not desirable in the online fashion. We are aiming to adapt our approach to update the model in a much faster way.

**Table of Alerts**

**Log Details**

attack from ip:10.0.0.2
req:/layer7/myweb/sample3/images/album_pics03.jpg
attempts:191

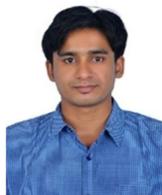

First Author: Prajwal R Thakare received his Bachelor of Engineering (BE) degree from Sant Gadge Baba Amravati University in 2011. He is currently pursuing M. Tech. in Information Technology, from ASTRA, Bandlaguda affiliated from Jawaharlal Nehru Technological University, Andhra Pradesh. His research interests are in Information and Network Security, Cloud Computing.

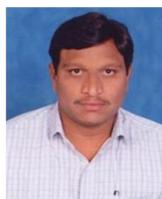

Second Author: K Hanumantha Rao, Associate Professor, Aurora Scientific Technological and Research Academy, completed his M. Tech (CSE). He published several research papers in the field of Network Security and Computer Networks. His areas of research are Network Security, Computer Networks and Cloud Computing.